\documentclass[twocolumn,english,aip,jcp,reprint]{revtex4-1}
\usepackage[T1]{fontenc}
\usepackage[latin9]{inputenc}
\usepackage{xcolor}
\usepackage{pdfcolmk}
\usepackage{refstyle}
\usepackage{rotfloat}
\usepackage{amsmath}
\usepackage{graphicx}
\PassOptionsToPackage{normalem}{ulem}
\usepackage{ulem}

\makeatletter

\AtBeginDocument{\providecommand\eqref[1]{\ref{eq:#1}}}
\AtBeginDocument{\providecommand\figref[1]{\ref{fig:#1}}}
\providecommand{\tabularnewline}{\\}
\RS@ifundefined{subsecref}
  {\newref{subsec}{name = \RSsectxt}}
  {}
\RS@ifundefined{thmref}
  {\def\RSthmtxt{theorem~}\newref{thm}{name = \RSthmtxt}}
  {}
\RS@ifundefined{lemref}
  {\def\RSlemtxt{lemma~}\newref{lem}{name = \RSlemtxt}}
  {}

\providecolor{lyxadded}{rgb}{0,0,1}
\providecolor{lyxdeleted}{rgb}{1,0,0}

\DeclareRobustCommand{\lyxsout}[1]{\ifx\\#1\else\sout{#1}\fi}

\usepackage{braket}
\usepackage[version=3]{mhchem}
\usepackage{url}

\@ifundefined{showcaptionsetup}{}{%
 \PassOptionsToPackage{caption=false}{subfig}}
\usepackage{subfig}
\makeatother

\usepackage{babel}
\begin{document}
\title{An assessment of initial guesses for self-consistent field calculations.
Superposition of Atomic Potentials: simple yet efficient}
\author{Susi Lehtola}
\affiliation{Department of Chemistry, P.O. Box 55 (A. I. Virtasen aukio 1), FI-00014
University of Helsinki, Finland}
\email{susi.lehtola@alumni.helsinki.fi}

\begin{abstract}
Electronic structure calculations, such as in the Hartree--Fock or
Kohn--Sham density functional approach, require an initial guess
for the molecular orbitals. The quality of the initial guess has a
significant impact on the speed of convergence of the self-consistent
field (SCF) procedure. Popular choices for the initial guess include
the one-electron guess from the core Hamiltonian, the extended Hückel
method, and the superposition of atomic densities (SAD).

Here, we discuss alternative guesses obtained from the superposition
of atomic potentials (SAP), which is easily implementable even in
real-space calculations. We also discuss a variant of SAD which produces
guess orbitals by purification of the density matrix that could also
be used in real-space calculations, as well as a parameter-free variant
of the extended Hückel method, which resembles the SAP method and
is easy to implement on top of existing SAD infrastructure.

The performance of the core Hamiltonian, the SAD and the SAP guesses
as well as the extended Hückel variant is assessed in non-relativistic
calculations on a dataset of 259 molecules ranging from the first
to the fourth periods by projecting the guess orbitals onto precomputed,
converged SCF solutions in single- to triple-$\zeta$ basis sets.
It is shown that the proposed SAP guess is the best guess on average.
The extended Hückel guess offers a good alternative, with less scatter
in accuracy.
\end{abstract}
\maketitle
\global\long\def\ERI#1#2{(#1|#2)}%
\global\long\def\bra#1{\Bra{#1}}%
\global\long\def\ket#1{\Ket{#1}}%
\global\long\def\braket#1{\Braket{#1}}%

\newcommand*\citeref[1]{ref. \citenum{#1}} 
\newcommand*\Citeref[1]{Ref. \citenum{#1}} 
\newcommand*\citerefs[1]{refs. \citenum{#1}}
\newcommand*\smallbas{pcseg\mbox{-}1}
\newcommand*\bigbas{(aug\mbox{-})pcseg\mbox{-}2}

\section{Introduction\label{sec:Introduction}}

Quantum chemical calculations are used in several applications to
determine single-point energies or molecular properties of systems
of interest. The level of theory can range from mean-field Hartree--Fock
(HF) or Kohn--Sham (KS) density functional theory\citep{Hohenberg1964,Kohn1965}
(DFT) to high-level \emph{ab initio} methods, such as multiconfigurational
(MC) self-consistent field (SCF) theory,\citep{Hinze1973} coupled-cluster
(CC) theory,\citep{Cizek1966} or the density matrix renormalization
group (DMRG)method.\citep{White1992} In each of these approaches,
the energy can be written in terms of a reference set of orbitals.
Solving the electronic structure is then tantamount to minimizing
the energy with respect to the reference orbitals.

High-level \emph{ab initio} methods are invariably initialized with
HF or KS orbitals. As HF produces by definition the best possible
single-configurational wave function, it often offers a reasonable
starting point for the treatment of correlation effects. Conversely,
KS typically produces good orbitals in cases where HF is not a good
starting point, such as for transition metal complexes. Because of
this, for the present purpose it is sufficient to restrict the discussion
to the HF and KS levels of theory.Although the HF and KS theories
are mathematically simpler than high level \emph{ab initio }methods
such as MC-SCF or DMRG-SCF, the minimization of the corresponding
energy functional is still a difficult non-linear optimization problem
that has been tackled with dozens of robust methods that cannot be
satisfactorily reviewed here due to length constraints.

Regardless of the method used to optimize the orbitals, an initial
guess is necessary. Orbital optimization is usually the simpler, the
closer the initial guess is to the converged result. However, despite
its pronounced importance, the choice of initial orbitals has attracted
surprisingly little interest in the literature.\citep{Weinstein1968,Vacek1999,Amat2002,Buenker2002,VanLenthe2006,Szekeres2006,Lee2015,Lim2016}
(Note that although the optimization problem can also be reformulated
only in terms of density matrices in the case of HF and KS theory,\citep{Seeger1976}
this has no implications for the present study, as the two approaches
are equivalent.)

As the HF / KS potential is density dependent, the simplest sensible
orbital guess (ignoring the trivial random orbital guess) is obtained
by minimizing the density independent part of the functional. By employing
the variational principle, it can be seen that in an orthonormal basis
set $\left\{ \ket i\right\} $, this task is equivalent to finding
the lowest eigenpairs of the matrix of the core Hamiltonian
\begin{equation}
H_{\text{core}}=\braket{i|\hat{T}+\hat{V}_{\text{nuc}}|j},\label{eq:Hcore}
\end{equation}
where $\hat{T}$ is the (single-determinant) kinetic energy and $\hat{V}_{\text{nuc}}$
is the nuclear attraction operator. Correspondingly, the guess from
\eqref{Hcore} is known as the \emph{core} or \emph{one-electron guess}.
If only one nucleus is present in the system, \eqref{Hcore} is the
hydrogenic Hamiltonian, and the core guess yields hydrogenic orbitals.

Now, as the core guess neglects all interactions between the electrons,
it does not take into account the significant screening of the nuclear
charge by the core electrons, thereby wasting considerable effort
to converge the shell structure of atoms. Furthermore, as the initial
guess does not reproduce the true energy ordering of atomic orbitals
of $s$, $p$, $d$, and $f$ symmetry, the procedure may produce
a molecular guess that does not have the symmetry of the ground state
solution. This may lead to the SCF algorithm requiring many more iterations
to find the ground state, or possibly even to the SCF procedure converging
onto a higher lying solution or a saddle-point, as \emph{e.g.} we
have recently shown for fully numerical calculations on diatomic molecules.\citep{Lehtola2019b}

Even worse, when applied to large systems composed of a diversity
of elements, the core guess will tend to crowd the heaviest atoms
with a large surplus of electrons, leaving the other atoms in highly
ionized states, because the hydrogenic orbital energies scale as the
square of the atomic number, $\epsilon_{i}\propto-Z^{2}$, while the
number of occupied orbitals per atom only scales as $n_{\text{occ}}\propto Z$.
This is not all, however. Hydrogenic orbitals are a famously poor
choice for single atoms as well, as the orbitals quickly become too
diffuse, thereby missing the important structure in the core and valence
regions,\citep{Helgaker2000} further highlighting the significant
shortcomings of the core guess both near and far from the nucleus.
(Note that while hydrogenic orbitals do not form a complete basis
set and have to be supplemented with continuum orbitals in order to
achieve good results,\citep{Hylleraas1928,Shull1955} this is not
a problem for the core guess, as the eigendecomposition of \eqref{Hcore}
does not change the dimension of the basis.)

The generalized Wolfsberg--Helmholz (GWH) approximation \citep{Wolfsberg1952}
is used in \citeref{Shao2015} as an alternative guess to the core
Hamiltonian, and it is the default guess for open-shell systems in
\citeref{Parrish2017}. In the GWH guess, the off-diagonal elements
of the Hamiltonian are approximated as
\begin{equation}
H_{ij}=\frac{1}{2}K\left(H_{ii}+H_{jj}\right)S_{ij},\label{eq:GWH}
\end{equation}
where the parameter $K$ typically has the value $K=1.75$, $H_{ii}$
and $H_{jj}$ are diagonal elements of the core Hamiltonian, and $S_{ij}$
is the overlap of basis functions $i$ and $j$. Although in some
cases the GWH modification of the core guess yields better results
than the core guess itself, it no longer yields an exact solution
for one-electron systems.

All of the problems with the core guess and its GWH modification can
be avoided by the use of the superposition of atomic densities (SAD)
guess,\citep{Almlof1982,VanLenthe2006} which employs converged atomic
density matrices at each nucleus in the system. As SAD has the correct
shell structure, it typically reproduces orbital energy orderings
as well. Indeed, SAD is used as the default guess in most popular
quantum chemistry packages, such as \textsc{Gaussian},\citep{Gaussian16}
\textsc{Molpro},\citep{Werner2012} \textsc{Orca},\citep{Neese2012}
\textsc{Psi4},\citep{Parrish2017} \textsc{PySCF},\citep{Sun2018}
and \textsc{Q-Chem}.\citep{Shao2015} An underappreciated feature
of the SAD guess is that it allows for pursuing different charge states
for a system as well as ionic \emph{vs.} non-ionic solutions by manually
assigning charge states on the individual nuclei in the system; unfortunately,
this is only possible in some implementations.\citep{VanLenthe2006,erkale}
Thus, in most cases the SAD density matrix is charge neutral, meaning
it may not match the actual charge state of the system.

Furthermore, the density matrix produced by SAD is non-idempotent
and does not correspond to a single-determinant wave function, which
results in a non-variational energy. In fact, SAD yields a non-idempotent
density matrix even for single atoms, as the guess typically uses
either configuration-averaged densities\citep{Almlof1982,VanLenthe2006}
or calculations with fractionally occupied orbitals.\citep{Jansik2009}
In addition, the SAD density matrix is spin-restricted, meaning that
it may also represent a different spin state than the one targeted
in the calculation.

The solution to the problems caused by the non-idempotency and the
possibly incorrect spin and charge state of the SAD density matrix
is simple. In the procedure of \citeref{VanLenthe2006}, a spin-restricted
Fock matrix build is performed with the initial guess density, which
is then diagonalized to yield a set of guess orbitals, which are then
used to start the wanted type of calculation; this is the most commonly
used approach. (Some implementations of SAD use the non-idempotent
closed-shell density matrix for the first step of the SCF calculation,
reporting non-variational energies for the first iteration.) Instead
of a Fock build, guess orbitals can also be constructed from the Harris
functional\citep{Harris1985,Foulkes1989} that does not require the
guess density to be idempotent; this is the approach chosen by \textsc{Gaussian}\citep{Gaussian16}.

Alternatively to a Fock matrix build followed by diagonalization,
guess orbitals could also be obtained from a SAD guess by diagonalizing
its density matrix to obtain natural orbitals. We are not aware of
the this guess that we call SADNO having been explicitly considered
previously in the literature. SADNO has been available in \textsc{Erkale}\citep{erkale,Lehtola2012}
and \textsc{Q-Chem}\citep{Shao2015} for some time, implemented in
both programs by the present author. However, as SADNO arises spontaneously
in linear-scaling approaches that employ density matrix purification
methods,\citep{McWeeny1960,Millam1997,Palser1998,Challacombe1999,Niklasson2002a,Jordan2005,Kim2016}
it may have been used implicitly in previous work.

Another guess that has been widely used in the past is the extended
Hückel method.\citep{Hoffmann1963} In the extended Hückel method,
orbitals are obtained by diagonalizing an effective one-particle Hamiltonian,
the diagonal of which consists of approximate valence state ionization
potentials (IPs), $H_{ii}=-\text{IP}_{i}$, whereas the off-diagonal
is estimated using the GWH rule (\eqref{GWH}). Traditionally, a minimal
set of Slater functions is used as the basis set, which is often replaced
in Gaussian basis codes with STO-3G.\citep{Hehre1969} Semi-empirical
calculations such as the CNDO\citep{Pople1966} or INDO\citep{Pople1967}
models can also be used instead of the extended Hückel guess.

However, as the traditional formulation of the extended Hückel method,
like the CNDO and INDO models, only operates within a minimal valence
basis set, the accuracy of these three methods may thereby be quite
limited, which is presumably why they have been largely replaced with
the SAD guess. Still, an implementation of the extended Hückel method
for real-space calculations has been described recently with good
results.\citep{Lee2015} However, as \citeref{Lee2015} only considered
a SAD guess formed of exponential model atomic densities instead of
\emph{ab initio }atomic density matrices, it is possible that the
performance of the SAD guess was underestimated.

As the original formulation of the extended Hückel method only describes
valence orbitals, core orbitals were added in \citeref{Lee2015} by
inserting Slater orbitals with exponents estimated from Slater's screening
rules. However, instead of relying on pretabulated IPs and minimal
Slater orbital basis sets as in the original formulation of the extended
Hückel method, Norman and Jensen proposed a variant in \citeref{Norman2012}
in which the basis functions and the diagonal elements of the Hamiltonian
are adopted as pretabulated Gaussian expansions of occupied atomic
HF orbitals and their orbital energies, respectively. Note that this
approach is in line with Hoffmann's original proposal,\citep{Hoffmann1963}
since HF orbital energies are approximations to the ionization potential
according to Koopmans' rule.\citep{Koopmans1934}

In the present work, we employ an extension of Norman and Jensen's
approach for the extended Hückel guess. However, instead of relying
on pretabulated orbitals and orbital energies as Norman and Jensen
do, in the present work the atomic orbitals and orbital energies are
calculated directly in the used basis set, employing the same machinery
as is used for the SAD guess. The present Hückel approach is easy
to implement, completely parameter-free, requires no orbital projections,
and is directly applicable to both all-electron and effective core
potential calculations.

In addition to the aforementioned approaches, calculations may be
initialized recursively by reading in a converged density computed
in a smaller basis set. Such a procedure has recently been advocated
for real-space calculations;\citep{Lim2016} the use of confinement
potentials has also been found to help SCF convergence in the case
of extended real-space basis sets.\citep{Lehtovaara2009} In some
cases it is also possible to decompose the system into either single
molecules or chemically meaningful molecular fragments,\citep{Jansik2009,Jansik2009a,Zhang2016c}
and ``glue'' the orbitals together to form a good guess density
for the original calculation. However, the problem of the proper choice
of the guess orbitals is not solved by either of these approaches,
but rather just moved to the additional calculation(s) in the smaller
basis set, or delegated to the isolated molecular fragments.

Having reviewed existing methodologies, what else could be done? In
the present work, we study the Superposition of Atomic Potentials
(SAP) guess, which can be used within both atomic orbital as well
as real-space basis set approaches. We will describe how to generate
the atomic potentials used in the SAP guess for all of the chemically
relevant part of the periodic table, and how to implement the guess
efficiently in non-relativistic or scalar-relativistic molecular calculations.
The SAP guess is extensively benchmarked against the core Hamiltonian
guess and its GWH modification, two variants of the SAD guess (SAD
and SADNO) as well as the extended Hückel guess variant that were
outlined above. The non-relativistic benchmark calculations comprise
259 molecules consisting of $1^{\text{st}}$ to $4^{\text{th}}$ period
elements, employing a variety of basis sets.

The organization of the manuscript is the following. Next, in the
Theory section, we will present the theory behind the SAP approach.
Then, in the Methods section, we describe the benchmark dataset, the
SCF calculations, and the guess assessment. The Computational Details
section outlines how the atomic potentials were calculated, and how
the SAP potential can be efficiently implemented in molecular calculations
using non-relativistic or scalar-relativistic approaches. Then, in
the Results section, we will present extensive benchmarks of the core,
GWH, extended Hückel, SAD, SADNO, GSZ, and SAP guesses employing various
potentials. Finally, the article concludes in a brief Summary and
Discussion section. Atomic units are used throughout the manuscript.

\section{Theory\label{sec:Theory}}

As molecules are formed from atoms, in which a sometimes overwhelming
fraction of electrons -- the core states -- are but spectators in
chemistry, an atom-focused guess indeed makes the most sense, as it
is simple, and as it yields the correct zeroth order solution. As
such, there are two ways in which the aim of an atomic guess could
be realized.

First, the target could be to reproduce the atomic orbitals or the
atomic electron density itself, as is done in the SAD guess. In calculations
employing linear combination of atomic orbitals (LCAO) basis sets,
it is trivial to perform atomic calculations at the beginning of a
molecular calculation, because the atomic basis sets are small --
especially since no polarization functions are necessary in the atomic
calculations. However, on-the-fly atomic calculations are intractable
for molecular calculations employing real-space methods, as performing
the atomic calculation in the three-dimensional molecular basis set
is inefficient. Thus, pretabulated atomic orbitals should be used
instead, requiring projections of the $N$ atomic orbitals onto the
molecular grid, followed by a construction of the density matrix in
the real-space basis set. We wish to point out here that although
\citeref{Lee2015} argued that the SAD guess only solves half the
problem for real-space electronic structure calculations by yielding
only a guess density but no guess orbitals, the approach we have outlined
here can be used to produce suitable guess orbitals for real-space
calculations. Namely, after projection of the numerical atomic orbital
SAD guess, molecular orbitals could be obtained with the SADNO scheme
by employing \emph{e.g.} a pivoted Cholesky decomposition of the SAD
density matrix which is feasible even for large systems.\citep{Aquilante2006}

Second, an atomic guess could be reproduced from a \emph{potential}
that yields the correct atomic electron density. This kind of a guess
would be equivalent to SAD in the case of non-interacting closed-shell
atoms, as the ground state of the atomic potential by definition yields
the orbitals for the atom. However, the use of a superposition of
atomic potentials (SAP) in a system of \emph{interacting} atoms might
produce a \emph{better} guess than that produced by the SAD method,
because the guess density will already be guided by interatomic interactions.
This can be illustrated by the following argument. Given electron
densities $\{n_{A}(\boldsymbol{r})\}$ on atoms $\{A\}$ that generate
potentials $\{v_{A}(\boldsymbol{r})\}$, respectively, the total energy
of the total system is given by
\begin{align}
E[n]= & \int n(\boldsymbol{r})v[n](\boldsymbol{r}){\rm d}^{3}r\label{eq:Etot0}\\
= & \int\left(\sum_{A}n_{A}(\boldsymbol{r})\right)v\left[\sum_{B}n_{B}(\boldsymbol{r})\right](\boldsymbol{r})\label{eq:Etot}\\
\approx & \int\left(\sum_{A}n_{A}(\boldsymbol{r})\right)\sum_{B}v\left[n_{B}(\boldsymbol{r})\right](\boldsymbol{r}){\rm d}^{3}r\label{eq:Elin}\\
= & \sum_{A}E[n_{A}]+\sum_{A\neq B}\int n_{A}(\boldsymbol{r})v\left[n_{B}(\boldsymbol{r})\right](\boldsymbol{r}){\rm d}^{3}r,\label{eq:E-SAP}
\end{align}
where we have approximated going from \eqref{Etot} to \eqref{Elin}
that the potential is linear in the density, as is the case for the
Coulomb and exact exchange potentials. The SAD guess corresponds to
the separate minimizations of the terms in the first sum, whereas
the SAP guess minimizes the total energy including the interatomic
interactions, thus yielding an improved guess density. However, as
SAP neglects the non-linear terms in \eqref{Etot}, the SAP guess
may be worse than SAD if the SAP density deforms a lot from the atomic
limit.

Compared to the many versions of the SAD guess or alternatives such
as the extended Hückel guess, the SAP guess is exceedingly simple
to implement. First, the atomic potentials for the SAP guess can be
easily obtained from calculations near the basis set limit, as we
have done in the present work; this step does not need to be replicated,
as non-relativistic and scalar-relativistic atomic potentials for
$1\leq Z\leq102$ are available from the present author. Second, the
formation of the SAP potential at any point involves but a simple
summation over the tabulated atomic potentials, which can be truncated
within a finite range. As a similarly local potential is also used
in the simplest variant of DFT, \emph{i.e.} the local spin density
approximation (LDA), existing DFT programs can be easily tailored
for the formation of the SAP guess. The analogy to DFT further shows
that the SAP potential matrix can be formed in linear scaling time
in large systems.\citep{Stratmann1996,Scuseria1999} Like the SADNO
guess we have proposed above, the SAP guess should be especially powerful
for real-space implementations, as in addition to producing a suitably
close guess density, it can also be used to produce a starting guess
for the orbital eigenvectors, for instance by solving its eigenstates
in a small basis of numerical atomic orbitals, or by iterative refinement
of the SAP orbitals to finer meshes.

The SAP potential can be reformulated by replacing the bare nuclear
attraction potential in the Hamiltonian 
\begin{equation}
V(\boldsymbol{r})=-\sum_{A}\frac{Z_{A}}{r_{A}},\label{eq:Vbare}
\end{equation}
where the sum runs over all atoms $A$ in the system, with a screened
version
\begin{equation}
V^{\text{SAP}}(\boldsymbol{r})=-\sum_{A}\frac{Z_{A}^{\text{eff}}(r_{A})}{r_{A}},\label{eq:VZeff}
\end{equation}
where the effective charge in \eqref{VZeff} can be trivially obtained
from the radial potential $V^{\text{SAP}}(r)$ produced by an atomic
calculation as
\begin{equation}
Z_{A}^{\text{eff}}(r)=-rV_{A}^{\text{SAP}}(r).\label{eq:ZVeff}
\end{equation}
The representation of \eqref{ZVeff} is extremely appealing, as it
removes any possible divergences of the potential at the nucleus:
as a consequence of \eqref{ZVeff}, the numerical range of $Z^{\text{eff}}(r)$
is limited and the function is smooth, making it easy to represent
numerically on a grid. Indeed, the canonical numerical representation
for the orbitals in atomic electronic structure programs is\citep{Lehtola2019a}
\begin{equation}
\psi_{nlm}(\boldsymbol{r})=r^{-1}P_{nl}(r)Y_{l}^{m}(\hat{\boldsymbol{r}}),\label{eq:atorb}
\end{equation}
which leads to the use of potentials $rV(r)$ for the radial functions
$P_{nl}(r)$.

If $Z^{\text{eff}}(r)$ only described the classical Coulomb potential,
it would be a monotonically decreasing function, going from $Z^{\text{eff}}(0)=Z$
at the nucleus to $Z^{\text{eff}}(\infty)=0$. But, in order to be
exact for atoms, quantum mechanical effects need to be included in
$V_{A}^{\text{SAP}}(r)$, meaning that $Z^{\text{eff}}$ may not be
monotonic. However, the limit $Z^{\text{eff}}(0)=Z$ still holds even
in the presence of exchange and correlation, and the function is overall
decreasing.

The present SAP approach is not fully novel, as somewhat reminiscent
approaches have been suggested earlier in the literature. It was recognized
already in the 1930s by Zener and Slater that the core electrons screen
the nuclear charge non-negligibly, and better approximate wave functions
can be obtained by employing an effective, shell-dependent screened
nuclear charge\citep{Zener1930,Guillemin1930,Slater1930} 
\begin{equation}
Z_{n}^{\text{eff}}=Z-s_{n},\label{eq:Slater-1}
\end{equation}
where $s_{n}$ is the screening constant for the shell $n$. Next,
the use of a radially screened nuclear potential for obtaining approximate
atomic orbitals for phenomenological studies was studied by Green
and coworkers in the late 1960s and early 1970s.\citep{Green1969,Green1973}
In contrast to Zener and Slater's rules, the approach used by Green
\emph{et al.} only has an \emph{implicit} shell dependence through
the Schrödinger equation: shells with $l>0$ experience a smaller
nuclear charge, since the $l(l+1)/r^{2}$ centripetal term\citep{Lehtola2019a}
in the kinetic energy prevents them from seeing the less-screened
regions close to the nucleus; this is also true for SAP. The Green--Sellin--Zachor
(GSZ) expression for the screened nuclear charge is given by\citealp{Green1969}
\begin{align}
Z^{\text{GSZ}}(r)= & 1+\frac{Z-1}{1+\left(e^{r/d_{Z}}-1\right)H_{Z}},\label{eq:zeff-gsz}\\
H_{Z}= & d_{Z}\left(Z-1\right)^{0.4}\label{eq:H-gsz}
\end{align}
where $d_{Z}$ is a nucleus specific parameter. Values for $d_{Z}$
for $Z\in[2,103]$ have been fitted to non-relativistic HF orbital
energies;\citealp{Green1969,Green1973} hydrogen is unaffected by
\eqref{zeff-gsz}. Slightly more refined GSZ-type potentials in which
also $H_{Z}$ is a free parameter have been obtained by minimization
of the HF energy of the wave function produced by the guess, reproducing
good agreement with numerical HF energies;\citep{Bass1973,Szydlik1974,Garvey1975,Green1975}
unfortunately, these potentials are not available for all the chemically
relevant parts of the periodic table.

Although Green and coworkers found the orbitals produced by the GSZ
potential to be close to the converged HF solutions, yielding good
agreement with experiment both for atoms\citep{Green1969,Ganas1971,Berg1971,Green1973,Green1973a,Bass1973,Riewe1973,Wallace1973,Szydlik1974,Garvey1975,Green1975,Doda1975,Green1976,Green1977}
as well as diatomic molecules,\citep{Whalen1972,Miller1974,Sawada1974}
the GSZ approach appears to be all but forgotten. The GSZ approach
\emph{is} available in the diatomic finite-difference HF program,\citealp{Kobus2013,x2dhf}
while the use of GSZ-inspired potentials for optimized effective potential
calculations has been studied by Theophilou and Glushkov.\citep{Theophilou2005,Theophilou2006}

\begin{figure*}
\centering{}\subfloat[BP86 screened nucleus\label{fig:BP86-screened-nucleus}]{\centering{}\includegraphics[width=0.33\textwidth]{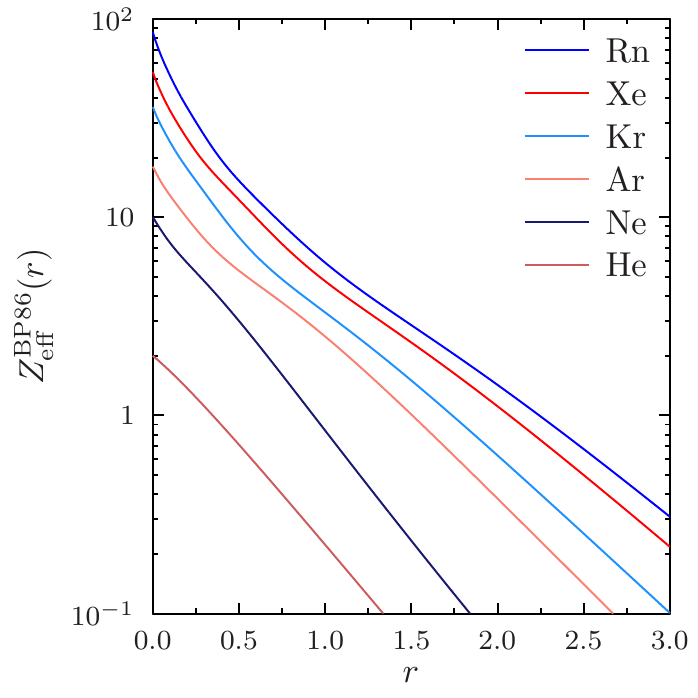}}\subfloat[BP86 atomic potential\label{fig:BP86-atomic-potential}]{\begin{centering}
\includegraphics[width=0.33\textwidth]{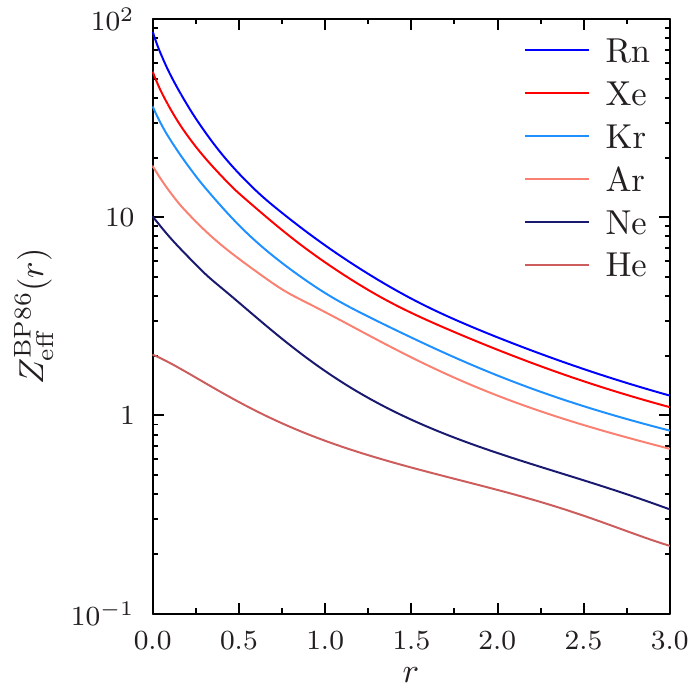}
\par\end{centering}
}\subfloat[GSZ\label{fig:GSZ}]{\begin{centering}
\includegraphics[width=0.33\textwidth]{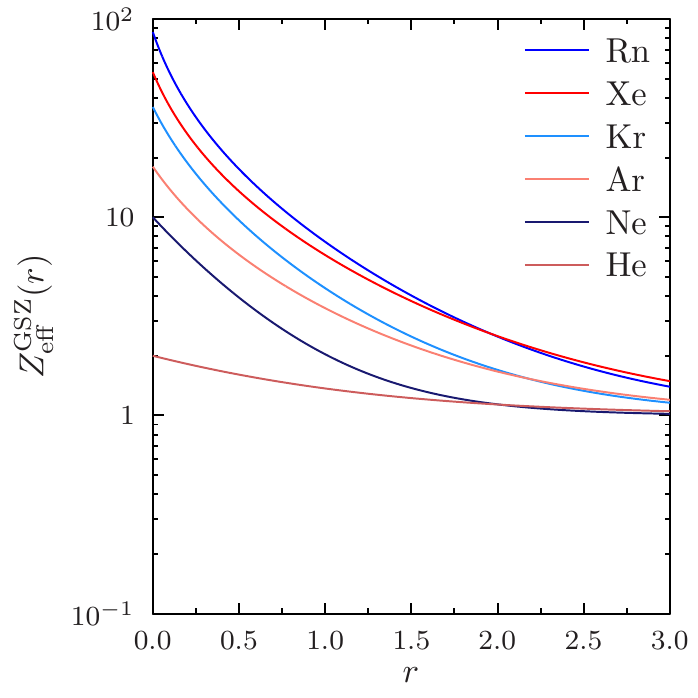}
\par\end{centering}
}\caption{Effective charges for the noble gas atoms, computed using the BP86
functional\citep{Becke1988a,Perdew1986} or given by the GSZ approach\citealp{Green1969}
(\eqref{zeff-gsz}). Note logarithmic scale. \label{fig:Effective-charge-for}}
\end{figure*}

At long range, the GSZ potential (\eqref{zeff-gsz}) has the asymptotic
value $Z^{\text{GSZ}}(\infty)=1$. Equivalently, far away, any SAP
potential should behave as $-1/r$. However, approximate exchange-correlation
potentials have an incorrect asymptotic form: potentials derived from
exchange-correlation energies decay in an exponential fashion,\citep{Almbladh1985}
meaning that DFT potentials will yield $Z^{\text{eff}}(\infty)=0$.
This behavior is illustrated in \figref{Effective-charge-for}, which
shows the effective nuclear charges given by \eqref{ZVeff} for the
SAP approach from a non-relativistic spin-restricted calculation with
the BP86\citep{Becke1988a,Perdew1986} functional (see Computational
Details section), and for the GSZ approximation. Effective charges
are also shown in \figref{Effective-charge-for} for a Coulomb-only
screened nucleus, based on the converged BP86 electron density. Comparison
of \figref{BP86-screened-nucleus, BP86-atomic-potential} shows that
while the classical screening of the Coulomb charge results in a rapid
decay of the effective charge, the inclusion of exchange and correlation
effects makes the atom much more attractive at chemically relevant
distances.

More recently, apparently unaware of the work by Green \emph{et al},
Amat and Cardó-Dorca\citep{Amat2002} suggested building guess orbitals
from an effective HF potential given by a prefitted,\citep{Amat2000}
spherically symmetric density representing the atomic shell structure,
motivated by the so-called atomic shell approximation (ASA).\citep{Constans1995}
(A similar approach for building guess orbitals from extended Hückel
calculations was suggested by Norman and Jensen.\citep{Norman2012})
ASA potentials have been fit for H--Ar\citep{Amat2000} and Sc--Kr\citep{Amat2002}
in the 6-311G basis set. In turn, the DIRAC program\citep{Dirac2017}
has employed a screened nuclear charge expressed as a Gaussian expansion
constructed from Zener and Slater's rules to obtain more accurate
guess orbitals for systems containing heavy atoms ever since its first
release in 2004.

Another approach was suggested by Nazari and Whitten,\citep{Nazari2017}
who optimized effective Gaussian potentials for H, C, N, O, and F
using a set of six molecules. The potentials were then benchmarked
for a test set of 20 molecules. Nazari and Whitten's results were
promising, showing that the model potentials are transferable between
different molecules to some extent. The method has been recently extended
to Ti, Fe, and Ni, as well as functional group specific potentials.\citep{Whitten2018}
However, as it is well known that orbital optimization in HF and DFT
can be reformulated as a problem of finding the right optimized effective
potential,\citep{Levy1982} it is difficult to estimate how well and
how easily the results of \citerefs{Nazari2017} and \citenum{Whitten2018}
can be generalized to the rest of the periodic table, or even how
the method generalizes beyond the non-standard\citep{Whitten2017}
``double-zeta'' and ``triple-zeta'' basis sets of ``near Hartree--Fock
atomic orbitals'' used in the study.

As machine learning will likely soon be able to predict accurate electron
densities in a cost-efficient fashion,\citep{Brockherde2017} it will
thereby likely also yield excellent initial guesses for SCF calculations.
In the mean time, the present work yields suitably accurate starting
guesses for the whole of the periodic table. The present work differs
significantly from those of Green and coworkers,\citep{Green1969}
Amat and Cardó-Dorca,\citep{Amat2002} and Nazari and Whitten,\citep{Nazari2017}
as
\begin{enumerate}
\item the form of our atomic potential is not restricted to a fixed analytic
form as in \citerefs{Green1969, Amat2002, Nazari2017}, but is instead
determined numerically in a tabulated form,
\item we employ \emph{unoptimized} atomic potentials constructed using fully
numerical calculations on atoms, not effective potentials optimized
for a molecular training set as in \citeref{Nazari2017} or potentials
fitted to reproduce atomic calculations in a specific basis set as
in \citeref{Green1969},
\item unlike \citerefs{Green1969, Amat2002, Nazari2017}, we present a set
of potentials for the whole chemically relevant periodic table (H--No
\emph{i.e.} $1\leq Z\leq102$), in both non-relativistic and scalar
relativistic variants, enabling practical calculations to be performed
on any system,
\item unlike \citerefs{Amat2002, Nazari2017}, the atomic potentials are
extensively benchmarked with calculations on entirely unbiased systems,
as even the basis sets used to generate the atomic potentials and
those used in the the molecular applications are fundamentally different.
\end{enumerate}

\section{Methods\label{sec:Methods}}

\subsection{Molecular dataset\label{subsec:Molecular-dataset}}

In the present work, we study the 183 non-multireference molecules
from the high-level W4-17 test set of first- and second-row molecules,\citep{Karton2017}
which we furthermore augment with a dataset composed of 50 transition
metal complexes from \citerefs{VanLenthe2006} and \citenum{Buhl2006}
(referred to as TMC), as well as 28 complexes containing $3^{\text{rd}}$
or $4^{\text{th}}$ period elements from the MOR41 database of single-reference
systems.\citealp{Dohm2018} Although the entries ED15, PR01, and PR02
in MOR41 are also included in TMC as \ce{Ni(C3H5)2}, \ce{Cr(CO)6},
and \ce{Fe(CO)5}, respectively, the existence of these duplicates
should not significantly affect our results as they represent only
a small fraction of the database, and the geometries for the molecules
are also slightly different. In contrast, \ce{Cr(C6H6)(CO)3}, \ce{CrO2F2},
\ce{Fe(CO)5}, \ce{VF5}, and \ce{VOF3} were excluded from \citeref{Buhl2006},
as these molecules also exist in \citeref{VanLenthe2006}. Moreover,
as only two molecules in the collection are charged, \ce{CrO4^{2-}}
and \ce{Co(NH3)6^{3+}} (both from \citeref{VanLenthe2006}), they
are omitted from the analysis due to insufficient representation.
The dataset of the present study thus contains 259 charge-neutral
molecules in total, 222 of which are singlets, and the remaining 37
are non-singlets.

\subsection{SCF calculations\label{subsec:SCF-calculations}}

Non-relativistic HF and revTPSSh\citep{Csonka2010,Perdew2009,Perdew2011}
wave functions were calculated for all the molecules using a development
version of \textsc{Q-Chem},\citep{Shao2015} employing wave function
stability analysis and a (99,590) integration grid for the exchange-correlation
functional. A $10^{-5}$ basis set linear dependence threshold, a
$10^{-12}$ integrals screening threshold, and a $10^{-6}$ SCF convergence
criterion was used. To investigate the impact of the basis set on
the performance of the guesses, calculations were performed in the
minimal STO-3G basis,\citep{Hehre1969} as well as the recently published
single- to triple-$\zeta$-level pcseg-0, pcseg-1, and aug-pcseg-2
basis sets.\citep{Jensen2014,Jensen2015} The motivation for this
range of basis sets is that the aug-pc-2 basis set\citep{Jensen2001,Jensen2002b}
that is the parent of aug-pcseg-2 has been recently found a good choice
for reasonably accurate calculations at the DFT level of theory.\citep{Mardirossian2017a}

\subsection{Guess assessment\label{subsec:Guess-assessment}}

The quality of a starting guess is determined by how close the orbitals
it yields are to the true ground state solution. Thus, the various
initial guesses are assessed in the present work by calculating the
projection of the guess orbitals onto the converged SCF wave function
as
\begin{equation}
Q^{\sigma}=\sum_{i,j=1}^{N_{\text{occ}}^{\sigma}}\left|\braket{i_{\text{\text{guess}}}^{\sigma}|j_{\text{\text{SCF}}}^{\sigma}}\right|^{2}=\text{Tr }\boldsymbol{P}_{\text{guess}}^{\sigma}\boldsymbol{S}\boldsymbol{P}_{\text{SCF}}^{\sigma}\boldsymbol{S},\label{eq:criterion}
\end{equation}
where $i$ and $j$ are molecular orbitals, the sums run over the
$N_{\text{occ}}^{\sigma}$ occupied orbitals of spin $\sigma$, $\boldsymbol{P}_{\text{guess}}^{\sigma}$
and $\boldsymbol{P}_{\text{SCF}}^{\sigma}$ are the guess and SCF
density matrices, and $\boldsymbol{S}$ is the overlap matrix. While
the examination of SCF convergence characteristics has been used \emph{e.g.
}in the study by van Lenthe \emph{et al},\citep{VanLenthe2006} it
is non-trivial to discern between the effects of the initial guess
and that of the SCF algorithm in such an approach. In contrast, the
projection $Q^{\sigma}$ yields an unambiguous appraisal of initial
guesses, $0\leq Q^{\sigma}\leq N_{\text{occ}}^{\sigma}$; it is also
continuous rather than discrete like the number of SCF iterations,
yielding a more fine-grained ranking. Furthermore, as only a single
SCF calculation is necessary on each system in a $Q^{\sigma}$ based
approach, it is possible to explore many kinds of initial guesses
cost-efficiently with the present methodology.

Because the SAD density matrix is non-idempotent, $Q^{\sigma}$ is
not a fully reliable estimator for the accuracy of the SAD guess.
As the largest natural orbital occupation numbers in SAD may be many
times larger than one, they may artificially inflate the value of
$Q^{\sigma}$ while representing an unphysical density. Moreover,
fractional occupation of the valence orbitals is a well-known trick
to aid SCF convergence;\citep{Rabuck1999,Kudin2002} in this respect
the non-idempotency of SAD can actually be helpful, although fractional
occupations can be formed for other guesses as well. Furthermore,
as the SAD density matrix is typically chosen spin-restricted and
charge neutral, the projections $Q^{\sigma}$ are not reliable estimates
of the resulting SCF convergence for non-singlet molecules, and/or
for charged molecules that were excluded from the present study due
to the scarsity of reference geometries. However, $Q^{\sigma}$ \emph{is}
a reliable estimator for the SADNO guess. As it extracts natural orbitals
from the SAD density matrix, the SADNO guess is able to form idempotent
density matrices, as well as to adapt to charged as well as spin-polarized
systems. As will be seen below, the SADNO guess consistently yields
better $Q^{\sigma}$ values than SAD.

To better be able to compare the performance of the guesses, the projections
in both spin channels are condensed into a single criterion. Because
it is clear that the number of electrons missed by a guess will scale
proportionally to system size, the criterion should be intensive rather
than extensive, lest the largest systems dominate the analysis entirely.
Thus, we choose the fraction $f$ of electron density covered, (worst)
$0\leq f\leq1$ (best),
\begin{equation}
f=\frac{\sum_{\sigma}Q^{\sigma}}{\sum_{\sigma}N_{\text{occ}}^{\sigma}},\label{eq:rel-proj-1}
\end{equation}
as the guess ranking criterion.

The various initial guesses are formed and assessed with the freely
available \textsc{Erkale} program\citep{Lehtola2012,erkale} by reading
in the basis sets and SCF wave functions from the \textsc{Q-Chem}
output. The following guesses are studied:
\begin{enumerate}
\item The core Hamiltonian guess, denoted as CORE.\label{enu:The-core-Hamiltonian}
\item The GWH guess, \emph{i.e.} the GWH modification of the core Hamiltonian.\label{enu:The-GWH-guess.}
\item The SAD guess. In the present work, the atomic densities are formed
in \textsc{Erkale} with HF employing spin-averaged, fractional orbital
occupations of the valence shells.\label{enu:The-SAD-guess.}
\item The extended Hückel guess, denoted as HUCKEL, with the atomic orbitals
and eigenvalues taken from calculations analogous to the ones in the
SAD guess.\label{enu:The-extended-H=0000FCckel}
\item The GSZ potential.\citep{Green1969}\label{enu:The-GSZ-potential.}
\item The SADNO guess, where the orbitals are obtained by diagonalizing
the SAD density matrix.\label{enu:The-SADNO-guess.}
\item The SAP guess, with the LDA-X, CAP-X and CHA-X potentials, described
below in the ``Generation of atomic potentials'' section.\label{enu:The-SAP-guess}
\end{enumerate}
As was mentioned in the Introduction, guesses \ref{enu:The-core-Hamiltonian}--\ref{enu:The-SAD-guess.}
are commonly used approaches, whereas
\begin{itemize}
\item guess \ref{enu:The-extended-H=0000FCckel} employs a parameter-free,
easily implementable variant of the extended Hückel guess that has
not been previously considered to our knowledge,
\item guess \ref{enu:The-GSZ-potential.} has not been previously considered
beyond diatomic molecules, and
\item guesses \ref{enu:The-SADNO-guess.} and \ref{enu:The-SAP-guess} have
not been previously considered at all in the literature.
\end{itemize}

\section{Computational details\label{sec:Computational-details}}

\subsection{Generation of atomic potentials\label{subsec:Generation-of-atomic}}

The atomic potentials are generated by KS-DFT calculations with an
all-electron atomic program employing spherical symmetry (\emph{i.e.}
fractional occupations) that is available as a part of the \textsc{Gpaw}
program package\citep{Larsen2009,Enkovaara2010} and which uses the
\textsc{Libxc} library\citep{Lehtola2018} to evaluate the exchange-correlation
functionals. The atomic program produces self-consistent solutions
of the radial Kohn--Sham equations. The atomic calculation is initialized
with a solution in a large Gaussian basis set, after which the solution
is further refined by a finite difference calculation on a radial
grid. With a simple modification, the atomic program was made to save
the converged radial potentials to disk. Default settings for the
convergence criteria (density converged to $10^{-6}$) were employed,
while a radial grid two times larger than the default (4000 instead
of 2000 points) was used, with the practical infinity set at the default
value of 50 bohr.

Initial experimentation with various functionals in \textsc{Libxc}
revealed that the best results were obtained from exchange-only calculations,
and that the best exchange functionals were the local spin-density
approximation (LDA-X),\citep{Bloch1929,Dirac1930} the ``correct
asymptotic potential'' (CAP-X),\citep{Carmona-Espindola2015} as
well as the Chachiyo\citep{Chachiyo2017} (CHA-X) generalized gradient
exchange functionals. Self-consistent atomic potentials were generated
for these three functionals at the non-relativistic and scalar-relativistic
levels of theory. The atomic calculations were spin-unrestricted,
and the SAP potential was generated by averaging over the two spin
channels of the converged potential. Next, as visual examination of
the potentials generated by GPAW revealed significant numerical noise
far away from the nucleus, the potentials were smoothed by forcing
them to decay exponentially far away from the nucleus as 
\begin{equation}
V(r)\to\begin{cases}
V(r), & r\leq r_{0}\\
V(r_{0})\exp(-\left[r-r_{0}\right]/\gamma) & r>r_{0}
\end{cases}\label{eq:Vdecay}
\end{equation}
with the onset $r_{0}=8a_{0}$ and the decay parameter $\gamma=4a_{0}$.
The GSZ potential\citep{Green1969} was implemented by pretabulating
its values in the same format as the potentials obtained from \textsc{Gpaw},
thus allowing maximal code reuse.

\subsection{Implementation of SAP\label{subsec:Implementation-of-SAP}}

The SAP potential matrix 
\begin{equation}
V_{\mu\nu}^{\text{SAP}}=\sum_{A}\int\chi_{\mu}(\boldsymbol{r})V_{A}^{\text{SAP}}(\boldsymbol{r})\chi_{\nu}(\boldsymbol{r}){\rm d}^{3}r,\label{eq:Vsap}
\end{equation}
where $V_{A}^{\text{SAP}}(\boldsymbol{r})$ is the repulsive potential
at $\boldsymbol{r}$ arising from atom $A$ and the sum runs over
all the atoms in the system, is calculated in \textsc{Erkale} using
Becke's polyatomic integration scheme\citep{Becke1988}
\begin{equation}
V_{\mu\nu}^{\text{SAP}}=\sum_{B}\int_{B}\chi_{\mu}(\boldsymbol{r})w_{B}(\boldsymbol{r})\left[\sum_{A}V_{A}^{\text{SAP}}(\boldsymbol{r})\right]\chi_{\nu}(\boldsymbol{r}){\rm d}^{3}r,\label{eq:Vsap-becke}
\end{equation}
which kills off any possible nuclear cusps in the potential. In analogy
to the SCF calculations, a (99,590) grid, \emph{i.e.} 99 radial and
590 angular points, is used for the SAP integrals (\eqref{Vsap-becke}).
The SAP potential in \eqref{Vsap-becke} is calculated using \eqref{VZeff},
in which linear interpolation is used for the pretabulated effective
charges $Z_{A}^{\text{SAP}}(r)$.

\section{Results\label{sec:Results}}

The statistics on the accuracy of the guesses described in the ``Guess
assessment'' section, assessed on HF/aug-pcseg-2 and HF/pcseg-0 wave
functions, is shown in \tabref{data}. The analysis in \tabref{data}
has been performed separately for the 222 neutral singlet molecules,
and for the 39 neutral non-singlet molecules of the present study.
\Tabref{data} also shows data for projections of wave functions calculated
at a different level of theory to study the accuracy of the commonly
used approach of reading in converged densities from another calculation.

As is shown by the large projection of the HF/aug-pcseg-2 and revTPSSh/aug-pcseg-2
wave functions, the level of theory is used to study the accuracy
of the initial guesses in a $Q^{\sigma}$ based approach does not
matter. After all, it is well known that HF and KS orbitals are typically
very similar (other than in pathological multireference cases such
as \ce{Cr2}); analogous results can be found in the literature.\citep{Stowasser1999,Kar2000}
We have thus shown that the level of theory used is not important
for the present approach: HF and KS references yield similar results.

Despite the resemblance of the orbitals at convergence, differences
in the SCF convergence characteristics of HF theory and DFT between
different initial guesses can likely be found. This is due to the
differing nature of the HF and KS potentials. The potential is orbital-dependent
in HF, whereas in DFT all orbitals experience the same potential;
however, the Taylor expansion of the KS energy is more complicated
than that of HF which is quadratic in the density. Thus, the differences
in the convergence speed of SCF calculations started from two guesses
of a similar accuracy will be dominated by the SCF acceleration technique,
of which there are many as stated in the beginning of the Introduction,
and which are known to behave differently even when started from the
same guess.

The results in \tabref{data} support the well-known procedure of
reading in an SCF solution from a smaller basis set: unsurprisingly,
a guess consisting of a converged SCF solution yields better results
than any of the \emph{ad hoc} guesses considered in the present work.
Based on its excellent coverage, we can recommend the single-$\zeta$
pcseg-0 basis as a guess basis for calculations in larger basis sets.
However, as discussed in the Introduction, an initial guess is still
necessary in the small basis set. Now, we continue by studying the
performance of the \emph{ad hoc} guesses in the aug-pcseg-2 and pcseg-0
basis sets.

The high quality of the SAP guess is demonstrated by the high $f$
values reproduced by all the three atomic potentials chosen for the
present work. The guess rankings in decreasing accuracy are in aug-pcseg-2
\begin{description}
\item [{Singlets}] mean $f$: CHA-X, CAP-X, LDA-X, HUCKEL, SADNO, GSZ,
SAD, CORE, and GWH; min $f$: HUCKEL, CAP-X, LDA-X, CHA-X, GSZ, SADNO,
SAD, CORE, and GWH.
\item [{Non-singlets}] mean $f$: CAP-X, CHA-X, LDA-X, HUCKEL, SADNO, GSZ,
SAD, CORE, and GWH; min $f$: CAP-X, CHA-X, LDA-X, HUCKEL, GSZ, (SAD,)
SADNO, CORE, and GWH.
\end{description}
and in pcseg-0
\begin{description}
\item [{Singlets}] mean $f$: CHA-X, CAP-X, HUCKEL, LDA-X, SADNO, GSZ,
SAD, CORE, GWH; min $f$: HUCKEL, CAP-X, LDA-X, CHA-X, SADNO, GSZ,
SAD, CORE, GWH.
\item [{Non-singlets}] mean $f$: CHA-X, LDA-X, CAP-X, HUCKEL, SADNO, GSZ,
(SAD,) CORE, GWH; min $f$: HUCKEL, CAP-X, SADNO, LDA-X, CHA-X, GSZ,
(SAD,) CORE, GWH.
\end{description}
This shows that on average, the SAP guess yields the best starting
point for calculations. The guess performance in the pcseg-0 and aug-pcseg-2
basis sets is also similar, underlining the quality of the proposed
approaches.

The extended Hückel variant described in the present work is also
a good performer, especially in the small pcseg-0 basis. The extended
Hückel guess can be seen as an approximation to SAP: in the version
described in the present manuscript, the atomic Hückel Hamiltonian
coincides with the atomic Fock operator that is diagonal in the Hückel
basis; note that the single-center atomic orbitals are orthonormal,
$S_{ij}=\delta_{ij}$. However, the Hückel guess approximates the
interatomic elements of the Hamiltonian with the generalized Wolfsberg--Helmholz
rule (\eqref{GWH}).

Furthermore, as the Hückel guess is limited to a minimal basis (even
though the basis functions themselves can constitute the exact solution
to the free atom), it yields a spectrum consisting mostly of zeros
for the virtual orbitals. In contrast, the SAP guess yields a full
spectrum for the virtual space also in a large basis set. However,
it is likely exactly the limitation to the minimal basis that allows
the Hückel guess to work so well: in contrast to the other \emph{ad
hoc} guesses considered here, the Hückel guess only mixes low-lying
states for the individual atoms, which means it can never stray very
far from a physical solution.

\begin{figure*}
\subfloat[GSZ]{\includegraphics[width=0.33\textwidth]{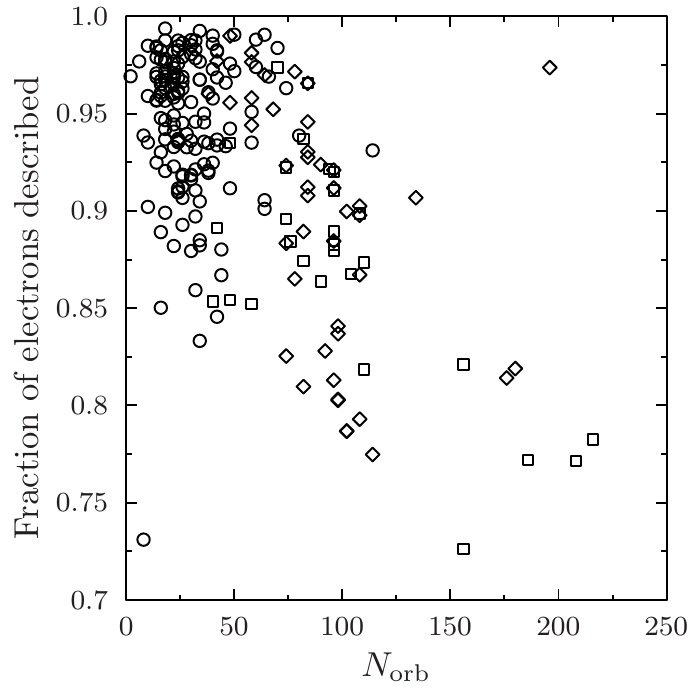}

}\subfloat[SADNO]{\includegraphics[width=0.33\textwidth]{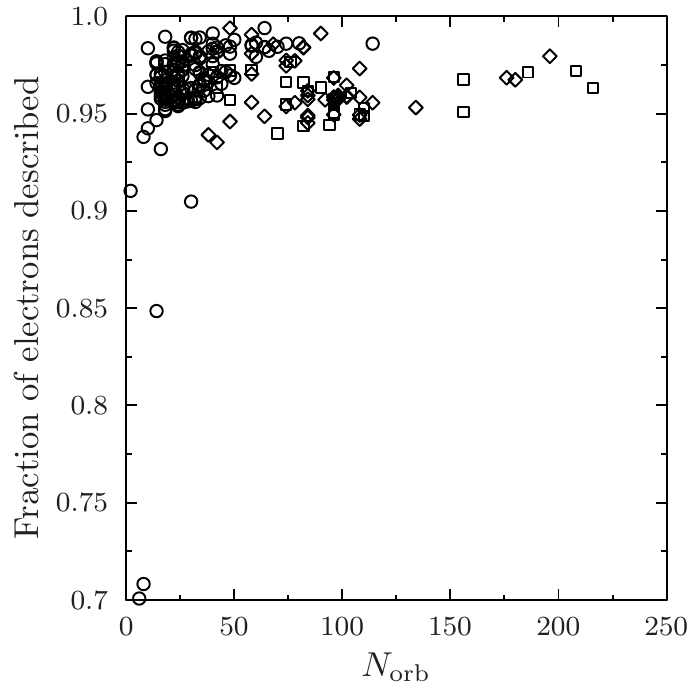}

}\subfloat[HUCKEL]{\includegraphics[width=0.33\textwidth]{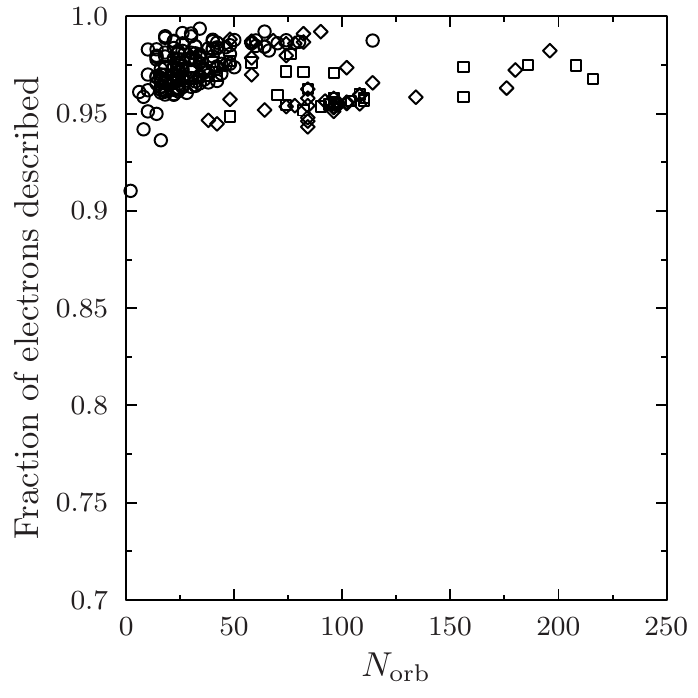}

}

\subfloat[LDA-X]{\includegraphics[width=0.33\textwidth]{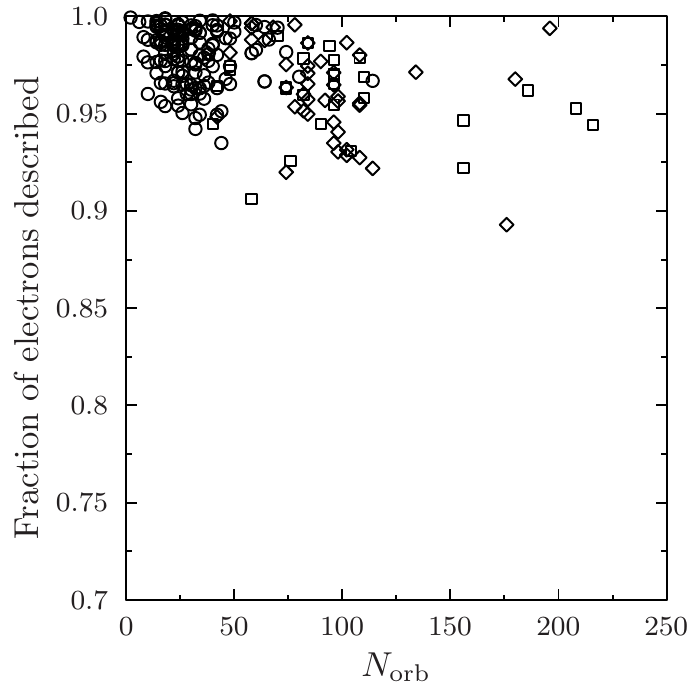}

}\subfloat[CAP-X]{\includegraphics[width=0.33\textwidth]{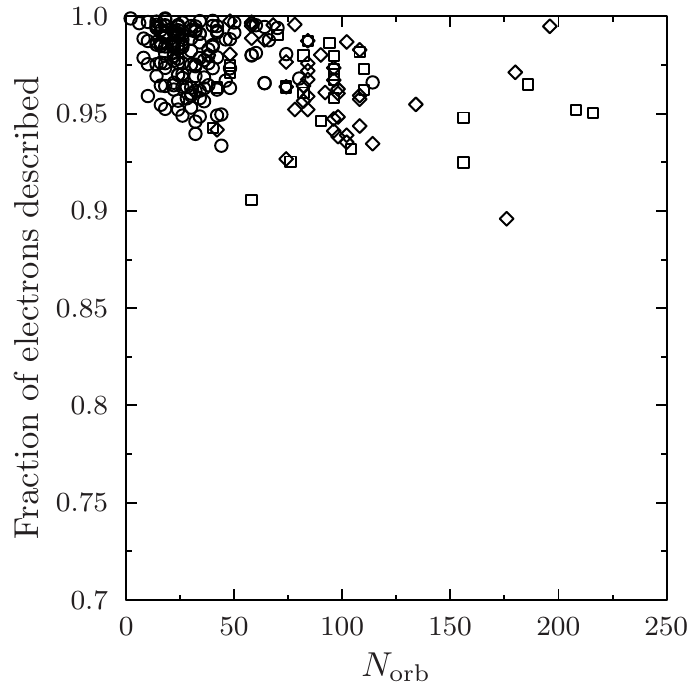}

}\subfloat[CHA-X]{\includegraphics[width=0.33\textwidth]{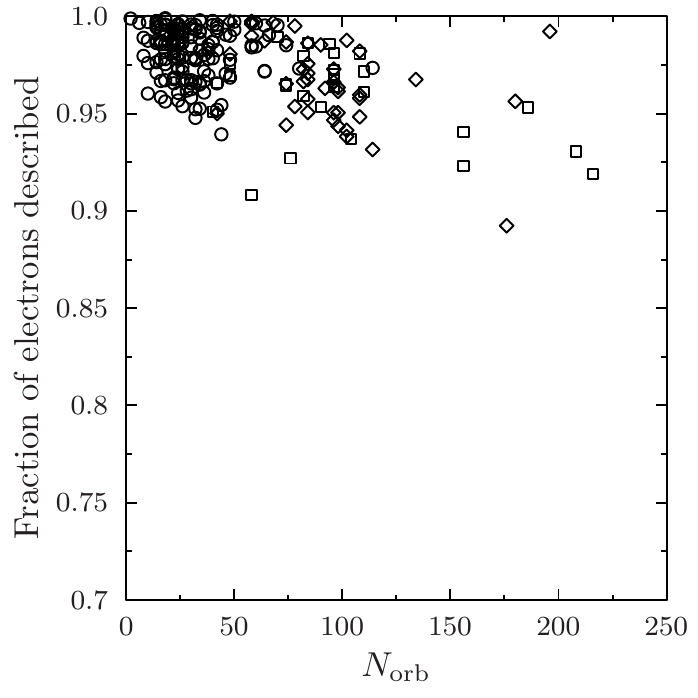}

}

\caption{Guess accuracy scatter plots for the HF/aug-pcseg-2 singlet wave functions.
Legend: non-multireference part of W4-17 (circles), transition metal
complexes from \citerefs{VanLenthe2006} and \citenum{Buhl2006} (diamonds),
MOR41 (squares).\label{fig:plots}}
\end{figure*}

The differences between the top performers can be studied in more
detail with the scatter plots shown in \figref{plots}. The GSZ potential
which has but a single parameter per atom exhibits a rapidly decaying
accuracy with increasing system size. The SAD guess, represented here
through the SADNO guess, generally offers a good starting point for
calculations, with some notable outliers in the case of small systems.
The extended Hückel variant is an improvement over SADNO, although
the scatter plots for the two methods share striking similarities;
after all, both guesses employ the same atomic calculations.

The three SAP methods are strikingly similar to each other. Although
the SAP results show considerably more scatter than the SADNO or the
extended Hückel guesses, SAP yields a more accurate initial guess
-- $f$ values closer to 1 -- for a large number of molecules.

\begin{sidewaystable*}
\begin{tabular}{lr@{\extracolsep{0pt}.}lr@{\extracolsep{0pt}.}lr@{\extracolsep{0pt}.}lr@{\extracolsep{0pt}.}l||r@{\extracolsep{0pt}.}lr@{\extracolsep{0pt}.}lr@{\extracolsep{0pt}.}lr@{\extracolsep{0pt}.}ll}
\textbf{aug-pcseg-2} & \multicolumn{4}{c}{singlets} & \multicolumn{4}{c||}{non-singlets} & \multicolumn{4}{c}{singlets} & \multicolumn{4}{c}{non-singlets} & \textbf{pcseg-0}\tabularnewline
Guess & \multicolumn{2}{c}{$\min f$} & \multicolumn{2}{c}{$\text{\text{mean} }f$} & \multicolumn{2}{c}{$\min f$} & \multicolumn{2}{c||}{$\text{\text{mean} }f$} & \multicolumn{2}{c}{$\min f$} & \multicolumn{2}{c}{$\text{\text{mean} }f$} & \multicolumn{2}{c}{$\min f$} & \multicolumn{2}{c}{$\text{\text{mean} }f$} & Guess\tabularnewline
\hline 
\hline 
GWH & 0&000 & 0&443 & 0&285 & 0&450 & 0&405 & 0&587 & 0&458 & 0&558 & GWH\tabularnewline
CORE & 0&435 & 0&585 & 0&417 & 0&610 & 0&523 & 0&680 & 0&557 & 0&662 & CORE\tabularnewline
SAD & 0&700 & 0&901 & 0&730 & 0&864 & 0&711 & 0&908 & 0&739 & 0&871 & SAD\tabularnewline
GSZ & 0&726 & 0&926 & 0&802 & 0&939 & 0&752 & 0&935 & 0&809 & 0&947 & GSZ\tabularnewline
SADNO & 0&701 & 0&964 & 0&715 & 0&946 & 0&758 & 0&973 & 0&861 & 0&959 & SADNO\tabularnewline
HUCKEL & 0&910 & 0&970 & 0&847 & 0&955 & 0&950 & 0&979 & 0&868 & 0&974 & HUCKEL\tabularnewline
LDA-X & 0&893 & 0&974 & 0&849 & 0&969 & 0&898 & 0&979 & 0&901 & 0&964 & LDA-X\tabularnewline
CAP-X & 0&896 & 0&974 & 0&898 & 0&973 & 0&901 & 0&979 & 0&851 & 0&975 & CAP-X\tabularnewline
CHA-X & 0&892 & 0&976 & 0&920 & 0&974 & 0&897 & 0&980 & 0&843 & 0&976 & CHA-X\tabularnewline
\hline 
\hline 
HF/STO-3G & 0&920 & 0&984 & 0&893 & 0&977 & 0&917 & 0&985 & 0&899 & 0&983 & HF/STO-3G\tabularnewline
HF/pcseg-0 & 0&932 & 0&996 & 0&929 & 0&991 & 0&974 & 0&998 & 0&976 & 0&997 & revTPSSh/pcseg-0\tabularnewline
revTPSSh/aug-pcseg-2 & 0&977 & 0&999 & 0&931 & 0&992 & \multicolumn{2}{c}{} & \multicolumn{2}{c}{} & \multicolumn{2}{c}{} & \multicolumn{2}{c}{} & \tabularnewline
HF/pcseg-1 & 0&963 & 0&999 & 0&994 & 0&999 & \multicolumn{2}{c}{} & \multicolumn{2}{c}{} & \multicolumn{2}{c}{} & \multicolumn{2}{c}{} & \tabularnewline
\end{tabular}

\caption{Statistics for the 222 neutral singlet molecules and the 37 neutral
non-singlet molecules at the HF/aug-pcseg-2 and HF/pcseg-0 levels
of theory.\label{tab:data}}
\end{sidewaystable*}

\section{Summary and Discussion\label{sec:Summary-and-Discussion}}

We have discussed an alternative method for obtaining an initial guess
for self-consistent field calculations that is based on the superposition
of atomic potentials (SAP), which is equivalent to the commonly used
superposition of atomic densities (SAD) approach in the case of systems
of non-interacting closed-shell atoms, for which both guesses are
exact. In the case of either open-shell atoms or molecules, neither
the SAP nor the SAD guess is exact. However, in contrast to SAD, the
guess formed by SAP also includes chemical interactions between the
atoms in a molecule in a linearized approximation.

The SAP approach can straightforwardly be implemented in programs
employing a linear combination of atomic orbitals, or a real-space
basis set. Once the SAP approach has been implemented, the choice
of the employed atomic potentials can be left to the user, greatly
facilitating future studies on better atomic potentials. Improvements
on the accuracy of the SAP approach could be obtained \emph{e.g. }by
manually specifying the charge states of the individual atoms in the
system, alike some implementations of SAD.

We have implemented the SAP approach in the freely available \textsc{Erkale}
program,\citep{Lehtola2012,erkale} which we have used for guess assessment
by projecting the guess orbitals onto the ground state wave functions
of a dataset consisting of 259 molecules comprised of $1^{\text{st}}$
to $4^{\text{th}}$ period elements. At variance to the assessment
of initial guesses by comparison of the resulting SCF convergence,
the results of which are highly dependent on the used SCF algorithm
and the assessment is further complicated by the possibility of convergence
to saddle point solutions or different minima, the presently used
projection approach yields an unambiguous accuracy score for any guess,
and also has a low computational cost that allows benchmarking a wide
variety of guesses.

In addition to SAP, we have discussed, implemented and assessed a
variant of SAD we call SADNO that produces guess orbitals by purification
of the non-idempotent SAD guess density matrix, which does not appear
to have been previously considered in the literature, as well as pointed
out and demonstrated that an extended Hückel guess can be easily implemented
on top of a pre-existing SAD solver, based on the procedure of \citeref{Norman2012}.

The SAP guess was shown to yield excellent guess wave functions in
combination with the Chachiyo generalized gradient exchange functional;\citep{Chachiyo2017}
almost as good results could also be obtained with the CAP\citep{Carmona-Espindola2015}
and LDA\citep{Bloch1929,Dirac1930} exchange functionals.

On average the SAP guess was best. However, there was more scatter
in the accuracy of SAP than in that of SADNO or the Hückel guess.
The accuracy of the SAP guess might be improved by forming the atomic
potentials at a better level of theory; for instance, effective potential
calculations,\citealp{Sharp1953,Talman1976} could be pursued in future
work.

The good results of the parameter-free extended Hückel guess variant
were explained through its connection to the SAP approach, as well
as through its minimal-basis structure that prevents it from yielding
very good or very bad performance. While its overall accuracy in the
present dataset was not as good as that of SAP, its accuracy is remarkably
stable. Because it is an improvement over SAD, and because it is extremely
easy to implement on top of pre-existing SAD code, we can recommend
the extended Hückel variant described in the present work as a default
choice.

While the present work considered only all-electron calculations at
the non-relativistic level of theory, the approaches discussed in
the present work are readily applicable to scalar-relativistic calculations,
and they also can be straightforwardly extended to calculations employing
effective core potentials. In the case of SAP, this would likely entail
the removal of the contributions from the core electrons to the SAP
potential, as the core electrons are already included in the effective
core potential.

The original motivation and driver of the present work was to develop
accurate yet easily implementable guesses for real-space approaches.\citep{Lehtola2019a,Lehtola2019b}
The SAP guess offers such an approach: suitable guess orbitals can
be easily obtained from a superposition of atomic potentials, which
are but simple scalar radial functions. Alternatively, the SAD guess
based on projection of pretabulated numerical orbitals could be used
to produce a guess density. If molecular orbital coefficients are
also needed, then the SADNO approach could be used to obtain them
from the SAD density. Finally, if pretabulated atomic orbitals and
orbital energies are already available for a SAD approach, the extended
Hückel variant studied in the present work following Norman and Jensen's
suggestion\citep{Norman2012} can also be easily implemented, again
yielding molecular orbital coefficients and a likely improved accuracy
over SAD.

\section*{Acknowledgments\label{sec:Acknowledgments}}

This work has been supported by the Academy of Finland through project
number 311149. Computational resources provided by CSC -- It Center
for Science Ltd (Espoo, Finland) and the Finnish Grid and Cloud Infrastructure
(persistent identifier urn:nbn:fi:research-infras-2016072533) are
gratefully acknowledged.

\bibliography{citations}

\end{document}